\documentclass[11pt]{article}

\usepackage[T1]{fontenc}
\usepackage[utf8]{inputenc}
\usepackage{microtype}
\usepackage{booktabs}
\usepackage{tabularx}
\usepackage{longtable}
\usepackage{array}
\usepackage{amsmath}
\usepackage{graphicx}
\usepackage{flafter}
\usepackage{placeins}
\usepackage{hyperref}
\usepackage[nameinlink,noabbrev]{cleveref}
\usepackage{url}
\usepackage{geometry}
\usepackage{caption}

\hypersetup{
    colorlinks=true,
    linkcolor=blue,
    citecolor=blue,
    urlcolor=blue,
    pdftitle={Mutation Testing for Reproducibility Safeguards in Machine Learning Research Software: An Empirical Study},
    pdfauthor={Ilya Shulepov},
    pdfsubject={Empirical study of mutation testing for reproducibility safeguards in machine-learning research software},
    pdfkeywords={mutation testing, reproducibility, machine learning, research software, software testing}
}

\title{
  Mutation Testing for Reproducibility Safeguards in\\
  Machine Learning Research Software: An Empirical Study
}

\author{
  Ilya Shulepov\\
  \small Independent Researcher\\
  \small \href{mailto:ia_shulepov1@student.mpgu.edu}
              {\texttt{ia\_shulepov1@student.mpgu.edu}}\\
  \small \href{https://orcid.org/0009-0001-1348-9576}
              {ORCID: 0009-0001-1348-9576}
}

\date{August 2026}

\begin{document}

\maketitle

\begin{abstract}
Reproducibility in machine-learning research depends on experimental choices
such as random seeds, dependency versions, data partitioning, and evaluation
configuration. Existing repository validation workflows may execute
successfully without detecting changes to such choices. We study this problem
using MLReproMutate, research software that applies controlled,
reproducibility-relevant mutations to ML research repositories and evaluates
them against validation workflows already present in those repositories.

We conducted an outcome-blind empirical study of 39 frozen
repository--operator cases using four mutation classes: random seed,
dependency pin, data split, and cross-validation fold count. Repository
revisions, mutation candidates, and validation workflows were fixed before
mutation outcomes were observed. Primary execution yielded outcomes for 13 of
39 cases; a bounded restoration procedure increased the combined evaluable set
to 24. After excluding one confirmed-equivalent mutation, 23 confirmed
non-equivalent mutations remained. The selected validation workflows detected
2 of these 23 mutations, corresponding to an observed detection proportion of
8.7\%.

These results do not imply that the corresponding repositories are
irreproducible. Rather, they show that, in this sample, existing validation
workflows often did not detect the particular controlled
reproducibility-relevant changes introduced by the study. The findings motivate
reproducibility-oriented mutation testing as a complementary way to assess
whether research-software safeguards constrain experimentally important
choices.
\end{abstract}

\section{Introduction}
\label{sec:introduction}

Reproducibility in machine-learning research depends not only on the
availability of source code, but also on experimental choices encoded in that
code and its execution environment. Random seeds, data partitioning,
cross-validation protocols, dependency versions, and related configuration
choices can affect measured results even when the software continues to execute
successfully \cite{reimers2017score,henderson2018matters,bouthillier2021variance}.
Accordingly, reproducibility initiatives increasingly emphasize the
publication of code, environments, experimental details, and executable
workflows \cite{gundersen2018reproducibility,pineau2021improving,heil2021standards}.

An open question is whether the validation mechanisms already present in
research repositories are sensitive to changes in these
reproducibility-relevant choices. A repository may contain tests, continuous
integration, validation scripts, documented experiments, or runnable examples.
Such workflows provide evidence that software can execute or that particular
properties hold, but they are not necessarily designed to detect changes in
experimental configuration.

This distinction is difficult to study from ordinary repository outcomes
alone. When a validation workflow succeeds, it is generally unknown whether it
would also succeed after a scientifically relevant experimental choice had
changed. Mutation testing provides a way to make this counterfactual explicit:
introduce a controlled change and observe whether the existing validation
workflow distinguishes the mutant from the original program
\cite{jia2011mutation,just2014mutants}.

Mutation testing has already been applied extensively to conventional software,
scientific software, and machine-learning systems
\cite{hook2009trustworthiness,ma2018deepmutation,humbatova2021deepcrime}.
Dependency-update testing has likewise used artificial faults to assess whether
test suites can detect problems introduced through changed dependencies
\cite{hejderup2022dependencies}. The present work therefore does not claim that
mutation testing itself is new to ML or research software. Instead, we use
mutation testing for a specific measurement target: the sensitivity of
\emph{existing validation workflows in real ML research repositories} to
controlled changes in reproducibility-relevant experimental choices.

We introduce MLReproMutate, research software that detects and applies
domain-specific mutations to reproducibility-relevant choices in Python ML
projects. In the present study, we consider four mutation classes:
\texttt{random-seed}, which changes a literal random seed;
\texttt{dependency-pin}, which relaxes an exact dependency constraint;
\texttt{data-split}, which disables explicit stratification in a
train/test split; and \texttt{cv-fold-count}, which changes an explicit
cross-validation fold count. Each mutation is evaluated against a validation
workflow selected from the repository rather than against a newly introduced
study-specific test oracle.

We use MLReproMutate in an empirical study of 39 frozen
repository--operator cases drawn from real ML research repositories. Corpus
construction was static and outcome-blind: repository eligibility, source
revision, mutation candidate, and validation workflow were fixed before
mutation outcomes were observed. For the primary B02 corpus, validation-oracle
categories were also recorded prospectively. Cases that could not execute at
baseline were kept separate from mutation outcomes rather than being replaced
after observing execution results.

Primary execution produced mutation outcomes for 13 of the 39 cases. A bounded
post-primary restoration procedure increased the combined evaluable set to 24
cases while preserving the frozen revision, mutation candidate, and validation
workflow. One evaluated mutation was confirmed semantically equivalent. Among
the remaining 23 confirmed non-equivalent mutations, the selected validation
workflows detected 2 and did not detect 21. Thus, \emph{in this sample}, the
observed confirmed mutation-detection proportion was 2/23 (8.7\%).

We additionally examine detection descriptively by validation-workflow type and,
for the prospectively classified B02 subset, by oracle type. The two detected
B02 mutations both occurred in the
\texttt{dependency-pin} $\times$ \texttt{completion-only} cell. Sparse cells,
unequal operator composition, and differential evaluability prevent attributing
this pattern to oracle strength. We therefore treat these comparisons as
descriptive rather than as evidence of causal or statistical differences among
workflow categories.

The study makes four contributions:

\begin{itemize}
    \item We present a reproducibility-oriented mutation model and its
    implementation in MLReproMutate, targeting experimental choices surrounding
    ML research workflows rather than learned-model structure alone.

    \item We define an outcome-blind empirical protocol that freezes repository
    revisions, mutation candidates, validation workflows, and prospective oracle
    metadata before mutation outcomes are observed.

    \item We provide empirical evidence from 39 frozen
    repository--operator cases, including explicit accounting for baseline
    executability, bounded restoration, and semantic equivalence.

    \item We distinguish mutation survival from repository-level
    reproducibility: a surviving confirmed non-equivalent mutation establishes
    only that the selected validation workflow did not detect that particular
    controlled change.
\end{itemize}

The remainder of the paper reviews related work
(Section~\ref{sec:related-work}), describes MLReproMutate and its mutation model
(Section~\ref{sec:mlrepromutate}), states the research questions and empirical
protocol, reports the results, and discusses their implications and threats to
validity.

\section{Background and Related Work}
\label{sec:related-work}

\subsection{Reproducibility and experimental variability in machine learning}
\label{sec:related-reproducibility}

Concerns about reproducibility in artificial intelligence and machine learning
span both research reporting and experimental practice. Gundersen and Kjensmo
surveyed empirical AI publications and found substantial gaps in the
documentation needed to reproduce reported experiments
\cite{gundersen2018reproducibility}. Raff subsequently studied independent
reimplementation of machine-learning research and emphasized that the
availability of code alone does not determine whether a result can be
independently reproduced \cite{raff2019reproducible}. Community interventions
such as the NeurIPS reproducibility program have therefore combined artifact
availability with reporting checklists and other changes to experimental
practice \cite{pineau2021improving}.

A related body of work has shown that apparently minor experimental choices can
materially affect measured ML performance. Reimers and Gurevych demonstrated
that random-seed choice can change reported results for neural sequence-tagging
systems \cite{reimers2017score}. Henderson et al.\ documented substantial
variance and reporting challenges in deep reinforcement learning
\cite{henderson2018matters}. Bouthillier et al.\ analyzed variance arising from
data sampling, parameter initialization, and hyperparameter selection in ML
benchmarks \cite{bouthillier2021variance}, while Pham et al.\ showed that both
algorithmic and implementation-level factors can introduce variance in deep
learning software \cite{pham2020variance}.

Reproducibility problems in ML-based science are not limited to stochastic
variation. Kapoor and Narayanan documented data leakage as a recurring source
of overoptimistic scientific conclusions \cite{kapoor2023leakage}. Heil et
al.\ similarly argue for explicit reproducibility standards spanning data,
models, code, programming practice, and workflow automation
\cite{heil2021standards}.

These studies motivate treating experimental configuration as part of the
scientific artifact rather than as incidental software detail. However, their
primary questions concern reproducibility, reporting, variance, or
methodological correctness. Our study asks a complementary testing question:
when a reproducibility-relevant experimental choice is deliberately changed,
does a validation workflow already present in the repository detect that
change?

\subsection{Executability and quality of research software}
\label{sec:related-executability}

Artifact availability is distinct from artifact executability. Collberg and
Proebsting reported substantial difficulties obtaining and rebuilding software
artifacts from computer-systems research \cite{collberg2016repeatability}.
At a larger scale, Trisovic et al.\ studied thousands of research-code files
from publicly available replication datasets and found that a large fraction
failed during clean-environment re-execution, even though some failures could
be mitigated through code cleaning \cite{trisovic2022execution}.

This distinction matters for mutation-based empirical studies. A mutation
cannot be meaningfully classified as detected or undetected when the
unmodified baseline cannot first execute under the selected validation
workflow. For this reason, our protocol treats setup, environment, and baseline
workflow failures as a separate evidence layer rather than assigning them
mutation outcomes.

The same literature also motivates preserving environment and workflow
information as part of research artifacts. Reproducibility recommendations for
ML emphasize explicit software environments, workflow automation, and
publication of the materials necessary to rerun an analysis
\cite{heil2021standards}. In our study, these concerns arise not only as
recommendations for future projects but also as practical constraints on
retrospective execution of historical repositories.

\subsection{Mutation testing and scientific software}
\label{sec:related-mutation}

Mutation testing evaluates a validation mechanism by introducing controlled
changes---mutants---and observing whether the available tests distinguish the
mutated program from the original. The technique has a long history in
software engineering; Jia and Harman provide a broad survey of its development,
operator design, cost, and mutation adequacy measures
\cite{jia2011mutation}.

A central justification for mutation analysis is that artificial faults can
provide a useful proxy for assessing test effectiveness when the complete set
of real faults is unknown. Just et al.\ compared mutant detection with
detection of real faults and found empirical evidence supporting this
relationship while also identifying limitations of the proxy
\cite{just2014mutants}. Mutation studies must also contend with mutants that do
not change relevant program semantics, motivating explicit treatment of
equivalent mutants rather than counting every syntactic change as an effective
fault.

Mutation techniques have also been considered in scientific software, where
the oracle problem can be particularly acute. Hook and Kelly used mutation in
the context of testing trustworthiness of scientific software, explicitly
addressing the difficulty of determining correct numerical outputs
\cite{hook2009trustworthiness}. This work is important precedent for applying
mutation concepts outside conventional business software.

Our use of mutation testing follows the same general adequacy principle but
changes the object being measured. We do not primarily ask whether a unit-test
suite detects arbitrary program faults. Instead, we use narrowly defined
reproducibility-relevant transformations to probe whether a frozen validation
workflow constrains experimental choices that may influence a research result.

\subsection{Mutation testing of machine-learning systems}
\label{sec:related-ml-mutation}

Mutation testing is already well established as a research direction for
machine-learning and deep-learning systems. DeepMutation introduced both
source-level mutations of training data and training programs and model-level
mutations of trained neural networks, using mutant detection to assess test-data
quality \cite{ma2018deepmutation}. DeepMutation++ extended mutation-based
evaluation for deep neural networks, including support for feed-forward and
recurrent architectures \cite{hu2019deepmutationpp}.

Subsequent work moved toward mutation operators grounded in observed DL faults.
DeepCrime derives source-level mutation operators from real fault evidence and
uses them to assess deep-learning test data
\cite{humbatova2021deepcrime}. DeepMetis in turn uses mutation score as an
optimization objective when augmenting DL test sets
\cite{riccio2021deepmetis}. These studies demonstrate that domain-specific
mutation operators can be more informative than simply applying conventional
program mutations to ML code.

There is therefore no novelty claim in this paper that mutation testing has not
previously been applied to ML systems. Some prior source-level DL mutation
operators also touch training and experimental configuration. The distinction
in our work is the measurement target and the oracle: MLReproMutate targets
reproducibility-relevant choices in research repositories and evaluates them
against validation workflows that existed in those repositories, rather than
primarily assessing the adequacy of model test data or the robustness of a
trained neural network.

\subsection{Dependencies, environments, and workflow reproducibility}
\label{sec:related-environments}

Dependency changes form an especially important overlap between reproducibility
engineering and mutation testing. A declared dependency modification is not
necessarily equivalent to a semantic environment change: package resolution
may select the same installed version, or a changed version may affect only
code that is not exercised by validation.

Hejderup and Gousios directly studied whether test suites can be trusted to
support automated dependency updates. Using Java projects and a large number of
artificial dependency faults, they found substantial gaps in the ability of
tests to detect semantic problems introduced through dependencies
\cite{hejderup2022dependencies}. Their work demonstrates that mutation-based
evaluation of dependency-update safeguards predates the present study.

Our \texttt{dependency-pin} operator addresses a different but adjacent
question in research software. It changes a repository's declared exact
constraint from \texttt{package==version} to
\texttt{package>=version}, resolves baseline and mutant environments
independently, and verifies whether the installed target dependency actually
changed before interpreting the validation result. The selected repository
workflow, rather than dependency-call coverage alone, serves as the empirical
oracle.

Environment concerns also connect dependency management to long-term
executability. Research-code studies show that software and environment drift
can prevent later execution \cite{trisovic2022execution}, while reproducibility
standards emphasize capturing and automating computational workflows
\cite{heil2021standards}. This motivates our separation between semantic
dependency changes, validation outcomes, and infrastructure failures.

\subsection{Positioning of this study}
\label{sec:related-positioning}

The literature above establishes substantial prior work on ML reproducibility,
research-code executability, conventional and scientific-software mutation
testing, mutation testing of deep-learning systems, and testing of dependency
updates. The contribution of the present study is therefore not a claim to be
the first use of mutation testing in ML, scientific software, or dependency
management.

Instead, the study combines these concerns around a specific empirical
measurement target: the sensitivity of existing ML research-repository
validation workflows to controlled changes in reproducibility-relevant
experimental choices. The empirical protocol operates on frozen revisions of
real research repositories, freezes mutation candidates and validation
workflows before observing outcomes, distinguishes infrastructure failures from
mutation outcomes, and explicitly verifies semantic equivalence where required.

This positioning also distinguishes mutation survival from a direct measure of
reproducibility. A surviving confirmed non-equivalent mutation shows that the
selected frozen workflow did not detect one controlled change. It does not
establish that the repository or its reported scientific result is
irreproducible. Conversely, a killed mutation demonstrates sensitivity to that
specific change rather than general reproducibility assurance.

\section{MLReproMutate and Mutation Model}
\label{sec:mlrepromutate}

MLReproMutate is Python research software for applying controlled
mutations to reproducibility-relevant choices in machine-learning projects. It
evaluates whether an existing repository validation workflow detects the
resulting change.

The tool differs from model-level mutation approaches that perturb learned
parameters, neural-network structure, or input examples. Its mutation targets
are instead elements of the surrounding experimental software and
configuration: choices such as random seeds, dependency constraints, data
partitioning, and evaluation-protocol parameters.

The present study uses four implemented operators. Throughout the paper we use
the concise study labels \texttt{random-seed}, \texttt{dependency-pin},
\texttt{data-split}, and \texttt{cv-fold-count}. These labels identify the
mutation classes used in the empirical protocol; the corresponding
implementation classes have more explicit software-level names.

\subsection{Operator interface and mutation candidates}
\label{sec:mlrepromutate-interface}

MLReproMutate represents a mutation operator through two conceptually separate
operations: candidate detection and mutation application. An operator first
identifies syntactically supported locations in a repository and represents
each location as a mutation candidate. A candidate records the operator,
reproducibility category, target file, a human-readable description, and
operator-specific metadata needed to identify and validate the target.

Mutation application is then performed against a selected candidate rather than
by searching for a new target at execution time. The implementation verifies
that the candidate still corresponds to the expected source location before
rewriting the file. This design reduces the risk that a mutation is silently
applied to a different syntactic occurrence after candidate selection.

For Python-source operators, candidate detection uses the Python abstract syntax
tree (AST) and records source-position metadata. The mutation is applied only
after the relevant call and literal or expression have been matched again
against the stored candidate information. The dependency operator similarly
records the target requirements file, package, version, and line position and
verifies the expected pin before replacement.

\subsection{Reproducibility mutation operators}
\label{sec:mlrepromutate-operators}

Table~\ref{tab:mutation-operators} summarizes the four operators used in the
study. The transformations are intentionally narrow: each operator recognizes
a restricted syntactic form and applies one deterministic mutation.

\begin{table}[t]
\centering
\caption{Reproducibility-relevant mutation operators used in the empirical
study. Each transformation is deterministic once a candidate has been
selected.}
\label{tab:mutation-operators}

\vspace{0.45em}

\small
\renewcommand{\arraystretch}{1.30}
\setlength{\tabcolsep}{7pt}

\begin{tabularx}{\textwidth}{
    >{\raggedright\arraybackslash}p{0.18\textwidth}
    >{\raggedright\arraybackslash}X
    >{\raggedright\arraybackslash}p{0.31\textwidth}
}
\toprule
\addlinespace[2pt]

\textbf{Study label} &
\textbf{Supported target} &
\textbf{Transformation} \\

\addlinespace[2pt]
\midrule
\addlinespace[4pt]

\texttt{random-seed} &
Literal integer argument to
\texttt{random.seed},
\texttt{np.random.seed},
\texttt{numpy.random.seed}, or
\texttt{torch.manual\_seed} &
$N \rightarrow N+1$ \\

\addlinespace[5pt]

\texttt{dependency-pin} &
Exact dependency pin
\texttt{package==version}
in a supported requirements file &
\texttt{package==version}
$\rightarrow$
\texttt{package>=version} \\

\addlinespace[5pt]

\texttt{data-split} &
Explicit non-\texttt{None}
\texttt{stratify} argument in
\texttt{train\_test\_split} &
\texttt{stratify=<expr.>}
$\rightarrow$
\texttt{stratify=None} \\

\addlinespace[5pt]

\texttt{cv-fold-count} &
Explicit literal \texttt{n\_splits} in
\texttt{KFold}, \texttt{StratifiedKFold},
\texttt{RepeatedKFold}, or
\texttt{RepeatedStratifiedKFold} &
$N \rightarrow N+1$
for \texttt{n\_splits} \\

\addlinespace[4pt]
\bottomrule
\end{tabularx}

\vspace{0.3em}
\end{table}

\paragraph{\texttt{random-seed}.}
The random-seed operator targets supported seed-setting calls whose first
positional argument is an integer literal. It does not attempt to resolve
variables, expressions, configuration files, or dynamically supplied seed
values. For an eligible literal seed $N$, the mutation replaces the literal
with $N+1$. The transformation preserves the presence of explicit seeding while
changing the selected pseudorandom sequence.

\paragraph{\texttt{dependency-pin}.}
The dependency-pin operator recognizes exact requirement specifications of the
form \texttt{package==version} and relaxes the equality constraint to
\texttt{package>=version}. The textual mutation alone does not guarantee that a
different package version will be installed. For resolved-dependency
evaluation, MLReproMutate therefore constructs baseline and mutant dependency
environments separately and records the installed version of the target
distribution. If both specifications resolve to the same installed version,
the dependency mutation is treated as equivalent rather than as evidence that
the validation workflow failed to detect a dependency change.

\paragraph{\texttt{data-split}.}
The data-split operator targets supported scikit-learn
\texttt{train\_test\_split} calls containing an explicit
\texttt{stratify} keyword whose value is not \texttt{None}. The operator
replaces the complete stratification expression with \texttt{None}. Candidate
detection resolves supported import forms before accepting a call, reducing
false matches to unrelated functions with the same local name.

\paragraph{\texttt{cv-fold-count}.}
The cv-fold-count operator targets supported scikit-learn cross-validation
splitters with an explicit literal integer \texttt{n\_splits} argument. The
supported splitter classes are \texttt{KFold},
\texttt{StratifiedKFold}, \texttt{RepeatedKFold}, and
\texttt{RepeatedStratifiedKFold}. For an eligible fold count $N \geq 2$, the
operator replaces it with $N+1$.

\subsection{Baseline-first mutation evaluation}
\label{sec:mlrepromutate-evaluation}

MLReproMutate evaluates mutations relative to an executable unmodified
baseline. Before evaluating selected candidates, the validation workflow is run
against an isolated copy of the unmodified project. A baseline timeout or
non-zero exit status prevents the corresponding mutation from being interpreted
as a validation outcome.

After a successful baseline, each mutation is applied in an isolated project
workspace and the same validation command is executed against the mutated
project. At the engine level, a successful validation command corresponds to a
SURVIVED mutation, a non-zero exit status corresponds to a KILLED mutation, and
a validation timeout is represented separately.

The implementation also defines additional internal states such as INVALID and
EQUIVALENT. In particular, resolved dependency evaluation can distinguish a
syntactically changed dependency specification that resolves to the same
installed version from one that actually changes the resolved environment.

The empirical protocol in Section~\ref{sec:study-design} imposes stricter
study-level rules on top of these software-level execution states. Setup and
baseline failures are kept separate from mutation outcomes, and semantic
verification determines which evaluated mutations enter the confirmed
non-equivalent detection denominator.

\subsection{Isolation and auditability}
\label{sec:mlrepromutate-auditability}

Mutation evaluation is performed in temporary project sandboxes rather than by
modifying the source repository in place. Candidate metadata and mutation
results provide an explicit link between the selected source location, applied
transformation, validation execution, and resulting outcome.

This separation is important for empirical use. Candidate detection can be
performed independently from mutation execution, selected candidates can be
frozen before outcomes are observed, and the same mutation definition can be
reapplied to an isolated repository copy. MLReproMutate therefore provides the
mutation mechanism and execution primitives, while corpus construction,
workflow selection, semantic verification, and the study-specific stopping
rules remain explicit parts of the empirical protocol described in
Section~\ref{sec:study-design}.

\section{Research Questions}
\label{sec:research-questions}

We investigate whether validation workflows already present in
machine-learning research repositories are sensitive to controlled changes in
reproducibility-relevant experimental choices. The study addresses two
research questions.

\paragraph{RQ1.}
When reproducibility-relevant mutations are introduced into ML research
software, how often are they detected by existing repository validation
workflows?

\paragraph{RQ2.}
How does mutation detection differ by validation-workflow type and by the
strength of the workflow's validation oracle?

RQ1 concerns the overall sensitivity of the selected validation workflows to
confirmed non-equivalent mutations. RQ2 examines this sensitivity
descriptively across workflow and oracle categories that were fixed
independently of mutation outcomes.

Here, oracle strength is operationalized through the prospectively recorded
oracle categories described in Section~\ref{sec:oracle-classification}; these
categories are not assumed to form an ordinal scale.

Neither research question treats mutation survival as evidence that a
repository is irreproducible. A surviving mutation indicates only that the
selected validation workflow did not detect that particular controlled change.

\section{Study Design}
\label{sec:study-design}

\subsection{Outcome-blind corpus construction}
\label{sec:corpus-construction}

The study used a frozen corpus of 39 repository--operator cases
drawn from machine-learning research software. Corpus construction was
performed separately from mutation execution and was intentionally
outcome-blind: repository eligibility, mutation candidates, revisions, and
validation workflows were selected without executing candidate repositories
or observing mutation outcomes.

The corpus consisted of two batches. B01 was a 10-repository calibration batch
used during development of the empirical protocol. B02 was the primary
operator-specific corpus and contained 29 repository--operator cases. The
final B02 composition was 10 \texttt{random-seed}, 10
\texttt{dependency-pin}, 6 \texttt{data-split}, and 3
\texttt{cv-fold-count} cases.

For B02, repositories were screened using an operator-specific static
eligibility procedure. Screening could use repository source code,
documentation, tests, continuous-integration configuration, experiment or
example source, and publication metadata, but could not execute repository
code, install dependencies, create environments, download datasets for
validation, or inspect mutation outcomes. In particular, repositories were
not selected because their tests appeared likely to kill a mutation or
because a candidate appeared especially sensitive to the proposed change.

The stopping rule was fixed before primary execution as

\[
n_{\mathrm{primary}} = \min(10, n_{\mathrm{eligible}}).
\]

Consequently, \texttt{random-seed} and \texttt{dependency-pin} each
contributed ten B02 cases, whereas the eligible static frames for
\texttt{data-split} and \texttt{cv-fold-count} were exhausted at six and
three cases, respectively. These smaller operator frames therefore represent
the eligible population identified under the frozen screening procedure
rather than post-hoc reductions based on execution outcomes.

The complete 39-case corpus, including repository revisions, candidate
mutations, and selected validation workflows, was frozen before primary
mutation execution. No repository was subsequently added to compensate for
setup failures, workflow failures, survived mutations, or other empirical
outcomes.

\subsection{Mutation operators and candidate selection}
\label{sec:mutation-operators}

The study used four reproducibility-relevant mutation operators implemented by
MLReproMutate. Each operator represents a controlled change to an experimental
or environment choice that can affect the reproducibility of an ML workflow.

\texttt{random-seed} targets supported literal calls to Python, NumPy, or
PyTorch random-seed APIs and changes an integer seed from \(N\) to \(N+1\).

\texttt{dependency-pin} relaxes an exact dependency pin. It replaces
\texttt{package}\allowbreak\texttt{==}\allowbreak\texttt{version} with
\texttt{package}\allowbreak\texttt{>=}\allowbreak\texttt{version}.

\texttt{data-split} removes a non-null \texttt{stratify} argument from a
supported \texttt{train\_test\_split} call.

\texttt{cv-fold-count} changes a literal \texttt{n\_splits=N} argument in a
supported KFold-family call to \texttt{n\_splits=N+1}.

Candidate selection was restricted to syntax supported by the frozen operator
definitions. The study therefore does not attempt to enumerate every
reproducibility-relevant choice that can occur in ML research software.
Instead, it evaluates a deliberately narrow set of mechanically identifiable
and auditable mutations.

For each selected case, the repository revision and candidate mutation were
fixed before execution. Candidate selection was not changed after observing
whether a baseline executed successfully or whether a mutant was detected.

\subsection{Frozen validation workflows}
\label{sec:validation-workflows}

Each case was associated with one validation workflow selected before mutation
execution. The objective was not to construct a new test suite for the
repository, but to measure the sensitivity of validation behavior that already
existed in the research artifact.

The protocol allowed five workflow categories:
\texttt{upstream-test}, \texttt{ci},
\texttt{documented-validation}, \texttt{documented-experiment}, and
\texttt{documented-example}. The selected command and its repository
documentation or source reference were frozen together with the case.

These categories describe the provenance or form of the workflow and are
treated categorically rather than ordinally. For example, an upstream test
command is not assumed a priori to provide a stronger reproducibility oracle
than a documented experiment, and an example workflow is not assumed to
provide a weaker one.

The term \emph{validation workflow} is intentionally broader than
\emph{reproducibility safeguard}. A workflow may execute substantial research
code while providing no explicit check on its outputs. Accordingly,
successful completion alone was not interpreted as evidence that a workflow
constituted a strong reproducibility oracle.

\subsection{Validation-oracle classification}
\label{sec:oracle-classification}

To distinguish workflow form from the mechanism that determines workflow
success, protocol version 2 prospectively classified the validation oracle for
every selected B02 case. Oracle classification was recorded before mutation
execution and was therefore independent of \textsc{Killed},
\textsc{Survived}, or \textsc{Equivalent} outcomes.

Four oracle categories were permitted. \texttt{assertion} indicates that
explicit assertions or test expectations determine success.
\texttt{metric-threshold} indicates an explicit numerical acceptance
criterion. \texttt{reference-comparison} indicates comparison with an expected
value or reference artifact. \texttt{completion-only} indicates that the
selected workflow is considered successful if it terminates with exit status
zero, without an explicit assertion, threshold, or reference comparison
relevant to the workflow result.

The frozen B02 corpus contained 22 \texttt{completion-only} and 7
\texttt{assertion} workflows; no B02 case used a
\texttt{metric-threshold} or \texttt{reference-comparison} oracle.

For a secondary descriptive contrast fixed before outcome analysis, the latter
three explicit-oracle categories were grouped as
\texttt{substantive-oracle}, while \texttt{completion-only} remained
separate. This binary grouping does not imply an ordering among
\texttt{assertion}, \texttt{metric-threshold}, and
\texttt{reference-comparison}.

B01 predates the protocol-v2 oracle field. We therefore retain the frozen B01
workflow categories but do not retrospectively treat B01 cases as if an oracle
category had been prospectively recorded. The primary oracle analysis for RQ2
is consequently restricted to B02.

\subsection{Primary execution protocol}
\label{sec:primary-execution}

Primary execution followed a baseline-first protocol. For each frozen case,
the selected repository revision was prepared according to the study protocol
and the frozen validation workflow was first executed on the unmodified
baseline. Mutation execution proceeded only when the baseline satisfied the
protocol's evaluability requirements.

If the baseline could not be established because of environment setup failure,
unavailable workflow requirements, or other infrastructure barriers, the case
was classified as non-evaluable at the primary layer rather than assigned a
mutation outcome. Mutation outcomes were therefore defined only for cases in
which the corresponding baseline and mutant could be evaluated under the
frozen workflow.

A mutation was classified as \textsc{Killed} when the selected validation
workflow failed or returned a non-zero status after the mutation. It was
classified as \textsc{Survived} when the same workflow completed successfully
despite the mutation. A mutation was classified as \textsc{Equivalent} when
semantic verification established that the applied mutation had not changed
the relevant semantics of the evaluated case.

Setup failures, workflow-unavailability events, and other infrastructure
failures were not counted as mutation outcomes.

\subsection{Semantic verification}
\label{sec:semantic-verification}

Mutation survival alone does not establish that a mutation changed the
semantics of the evaluated computation. The analysis therefore separates
validation-workflow detection from semantic equivalence.

Semantic verification distinguished four states: confirmed
non-equivalent, confirmed equivalent, unverified, and not run. Confirmed
equivalent mutations were excluded from the meaningful detection denominator.

The primary detection measure used for RQ1 and RQ2 is

\[
\frac{\mathrm{KILLED}}
     {\mathrm{KILLED} +
      \mathrm{confirmed\text{-}non\text{-}equivalent\ SURVIVED}}.
\]

This distinction is important for interpreting survival. A confirmed
non-equivalent \textsc{Survived} mutation means that the selected frozen
validation workflow did not detect that particular reproducibility-relevant
change. It does not imply that the repository as a whole is irreproducible.

\subsection{Bounded restoration of non-evaluable cases}
\label{sec:d027-restoration}

Primary execution revealed substantial environment and dependency drift in
historical ML research repositories. To distinguish mutation sensitivity from
failures caused by contemporary execution environments, we applied a separate
bounded restoration protocol, D027, to a predefined subset of initially
non-evaluable cases.

D027 was an additional evidence layer rather than a revision of the primary
study. Canonical primary outcomes and failure classifications were retained
unchanged. Restoration preserved the frozen repository revision, mutation
candidate, validation workflow, and oracle classification.

Restoration followed several constraints intended to prevent outcome-directed
repair. Source-level compatibility patches were prohibited; validation
workflows could not be skipped or weakened; the baseline had to be restored
before mutation evaluation; the restoration recipe was fixed before executing
the mutant; and baseline and mutant were evaluated using symmetric restoration
conditions. Restoration attempts were bounded, and infrastructure failures
were kept distinct from mutation outcomes.

The D027 cohort contained 24 cases. A structured D027 report or assessment was
available for 19 cases. Substantive restoration was attempted for 18 cases, 11
of which were successfully restored and subsequently evaluated. Seven cases
were not restored after substantive attempts. One additional case had a D027
report but no substantive restoration attempt because repository identity
recovery failed, and five cohort cases had no D027 report.

Successful D027 evaluations were combined with primary evaluations only at the
analysis layer. D027 never changed the frozen workflow or oracle
classification and never replaced the canonical primary record.

\subsection{Analysis procedure and denominators}
\label{sec:analysis-denominators}

We distinguish three levels of accounting: the complete frozen corpus, the set
of evaluable cases, and the set of confirmed non-equivalent mutations that
enter the meaningful detection denominator.

The complete corpus contains 39 cases. Primary execution evaluated 13 cases
and left 26 non-evaluable. After incorporating successful D027 restoration, 24
cases were evaluable and 15 remained non-evaluable.

Among the 24 combined evaluated cases, one mutation was confirmed equivalent.
The meaningful detection denominator therefore contains 23 confirmed
non-equivalent mutations: 2 \textsc{Killed} and 21 \textsc{Survived}.

RQ1 reports detection over these 23 cases. RQ2 first stratifies the same
combined meaningful cases by the frozen workflow category. The primary oracle
analysis is restricted to B02 because only B02 contains prospectively recorded
oracle classifications. Fifteen B02 cases entered the confirmed
non-equivalent meaningful denominator for this analysis: 2 \textsc{Killed}
and 13 \textsc{Survived}.

RQ2 is treated as descriptive and exploratory. Category sizes are small and
uneven, and only two confirmed non-equivalent mutations were detected. We
therefore report counts and descriptive proportions rather than interpreting
category differences as causal effects or evidence of statistical superiority.

As an additional diagnostic for RQ2 interpretation, we examined operator
composition within the prospectively classified B02 oracle groups and
evaluability attrition across categories. These diagnostics were performed
after the workflow and oracle classification frame had been independently
frozen. They are used to identify potential confounding and denominator
imbalance, not to redefine the RQ2 categories.

\section{Results}
\label{sec:results}

\subsection{Executability and bounded restoration}
\label{sec:results-executability}

Primary execution evaluated 13 of the 39 frozen cases (33.3\%), while 26
cases (66.7\%) were non-evaluable at the primary layer. These non-evaluable
cases were retained in the study frame rather than replaced by additional
repositories.

\begin{table}[t]
\centering
\caption{Executability before and after bounded restoration. Non-evaluable cases remain part of the frozen 39-case study frame.}
\label{tab:study-summary}
\vspace{0.35em}
\small
\renewcommand{\arraystretch}{1.28}
\setlength{\tabcolsep}{10pt}
\begin{tabular}{lrrr}
\toprule
\addlinespace[2pt]
\textbf{Layer} & \textbf{Selected} & \textbf{Evaluated, n (\%)} & \textbf{Non-evaluable, n (\%)} \\
\addlinespace[2pt]
\midrule
\addlinespace[3pt]
Primary execution & 39 & 13 (33.3\%) & 26 (66.7\%) \\
\addlinespace[3pt]
\textbf{Combined primary + D027} & 39 & \textbf{24 (61.5\%)} & \textbf{15 (38.5\%)} \\
\addlinespace[3pt]
\bottomrule
\end{tabular}
\vspace{0.25em}
\end{table}

The bounded D027 restoration layer increased the number of evaluable cases
from 13 to 24, corresponding to 61.5\% of the complete frozen frame. Fifteen
of 39 cases (38.5\%) remained non-evaluable after combining primary and
restoration evidence.

Within the 24-case D027 cohort, a report or assessment was present for 19
cases. Substantive restoration was attempted for 18 cases. Eleven cases were
successfully restored and evaluated, while seven were not restored after a
substantive attempt. One additional case had a D027 report but no substantive
restoration attempt, and five cohort cases had no D027 report.

These counts characterize executability under the study protocol rather than
mutation detection. In particular, setup and workflow failures were not
assigned KILLED or SURVIVED outcomes.

\FloatBarrier

\subsection{RQ1: Detection of reproducibility-relevant mutations}
\label{sec:results-rq1}

Among the 24 combined evaluated cases, one mutation was confirmed
semantically equivalent and was excluded from the meaningful detection
denominator. The resulting denominator contained 23 confirmed
non-equivalent mutations.

Of these 23 mutations, 2 were \textsc{Killed} and 21 were
\textsc{Survived}. The observed detection proportion was therefore

\[
\frac{2}{23} = 8.7\%.
\]

Thus, in this sample, the selected repository validation workflows detected
2 of 23 confirmed non-equivalent reproducibility-relevant mutations. The
remaining 21 confirmed non-equivalent mutations completed without detection
by their selected frozen workflows.

The B02 operator-specific results were uneven. All five meaningful
\texttt{random-seed} evaluations survived. For \texttt{dependency-pin},
four cases were evaluated: two were KILLED, one was SURVIVED, and one was
confirmed equivalent. Excluding the equivalent case yielded two detections
among three confirmed non-equivalent \texttt{dependency-pin} evaluations.
All four meaningful \texttt{data-split} cases and all three
\texttt{cv-fold-count} cases survived.

\begin{table}[t]
\centering
\caption{Combined B02 results by mutation operator. Detection is reported over confirmed non-equivalent evaluated mutations; confirmed-equivalent evaluations are excluded.}
\label{tab:operator-results}
\vspace{0.35em}
\small
\renewcommand{\arraystretch}{1.24}
\setlength{\tabcolsep}{6pt}
\begin{tabular}{lrrrrrr}
\toprule
\addlinespace[2pt]
\textbf{Operator} & \textbf{Eval./sel.} & \textbf{Non-eval.} & \textbf{Killed} & \textbf{Surv.} & \textbf{Eq.} & \textbf{Detection} \\
\addlinespace[2pt]
\midrule
\addlinespace[3pt]
\texttt{random-seed} & 5/10 & 5 & 0 & 5 & 0 & 0/5 (0.0\%) \\
\addlinespace[3pt]
\texttt{dependency-pin} & 4/10 & 6 & \textbf{2} & 1 & 1 & \textbf{2/3 (66.7\%)} \\
\addlinespace[3pt]
\texttt{data-split} & 4/6 & 2 & 0 & 4 & 0 & 0/4 (0.0\%) \\
\addlinespace[3pt]
\texttt{cv-fold-count} & 3/3 & 0 & 0 & 3 & 0 & 0/3 (0.0\%) \\
\addlinespace[3pt]
\bottomrule
\end{tabular}
\vspace{0.25em}
\end{table}

These operator-level counts are descriptive. The operator-specific
denominators are small and unequal, and the corpus construction was not
designed to estimate population-level differences in mutation detection
between operators.

\FloatBarrier

\subsection{RQ2: Detection by workflow and oracle type}
\label{sec:results-rq2}

We first stratified the 23 confirmed non-equivalent evaluated mutations by the
frozen validation-workflow category.

Among six \texttt{upstream-test} cases, none of the mutations was detected.
The single CI case also survived. One of two
\texttt{documented-validation} cases was detected, as was one of eleven
\texttt{documented-example} cases. None of the three
\texttt{documented-experiment} cases was detected.

\begin{table}[t]
\centering
\caption{Detection of confirmed non-equivalent mutations by frozen validation-workflow type. Workflow type is categorical; no ordinal ranking is assumed.}
\label{tab:rq2-workflow}
\begin{tabular}{lrrrr}
\toprule
Workflow type & N & Killed & Survived & Detection \\
\midrule
upstream-test & 6 & 0 & 6 & 0.0\% \\
CI & 1 & 0 & 1 & 0.0\% \\
documented-validation & 2 & 1 & 1 & 50.0\% \\
documented-experiment & 3 & 0 & 3 & 0.0\% \\
documented-example & 11 & 1 & 10 & 9.1\% \\
\bottomrule
\end{tabular}
\end{table}

The primary oracle analysis was restricted to B02 because oracle kind was
prospectively recorded only under protocol version 2. Fifteen B02 cases
entered the confirmed non-equivalent meaningful denominator. Of these, two
used an \texttt{assertion} oracle and thirteen used a
\texttt{completion-only} oracle.

Neither of the two \texttt{assertion} cases was detected. Two of the thirteen
\texttt{completion-only} cases were detected, corresponding to a descriptive
detection proportion of 15.4\%.

\begin{table}[t]
\centering
\caption{Detection by prospectively recorded validation-oracle kind in B02. B01 is excluded because schema version 1 did not prospectively record oracle kind.}
\label{tab:rq2-oracle}
\begin{tabular}{lrrrr}
\toprule
Oracle kind & N & Killed & Survived & Detection \\
\midrule
assertion & 2 & 0 & 2 & 0.0\% \\
completion-only & 13 & 2 & 11 & 15.4\% \\
\bottomrule
\end{tabular}
\end{table}

This pattern does not establish that completion-only workflows provide
stronger mutation detection than assertion-based workflows. A diagnostic
cross-classification by operator showed that both detected B02 mutations
occurred in the \texttt{dependency-pin} $\times$
\texttt{completion-only} cell: two of the three confirmed non-equivalent
cases in that cell were KILLED. No KILLED mutation occurred in any other
observed B02 operator--oracle cell.

The oracle categories also differed in evaluability. Of seven prospectively
classified \texttt{assertion} cases, three were combined-evaluable and two
entered the confirmed non-equivalent meaningful denominator. Of 22
\texttt{completion-only} cases, thirteen were combined-evaluable and all
thirteen entered the meaningful denominator.

Together, the sparse cells, operator composition, and differential
evaluability prevent attributing the observed detection pattern to oracle
strength. RQ2 therefore provides descriptive evidence of variation across
workflow and oracle categories, but no evidence of a monotonic relationship
between oracle explicitness and mutation detection.

\FloatBarrier

\section{Discussion}
\label{sec:discussion}

\subsection{Limited detection by existing validation workflows}
\label{sec:discussion-detection}

The central empirical observation of this study is that the selected validation
workflows detected 2 of 23 confirmed non-equivalent
reproducibility-relevant mutations. The remaining 21 mutations changed the
relevant semantics of the evaluated case but were not rejected by the frozen
validation workflow.

This result should be interpreted as a property of the evaluated
repository--mutation--workflow combinations rather than as a population-wide
estimate for ML research software. The corpus was constructed using a frozen,
operator-specific screening procedure, but the final meaningful denominator is
small and the operator-specific sample sizes are uneven. We therefore do not
infer that a fixed proportion of ML repositories would behave similarly under
other mutations or validation workflows.

Nevertheless, the observed survival of confirmed non-equivalent mutations
shows that successful execution of an existing repository workflow does not
necessarily imply sensitivity to reproducibility-relevant experimental
changes. A workflow may exercise the affected code path while still accepting
a changed seed, split, fold count, or resolved dependency environment.

The result is therefore better understood as evidence about \emph{validation
sensitivity}: in 21 of the 23 confirmed non-equivalent evaluations, the
selected existing workflow did not distinguish the original configuration from
the controlled, semantically different one.

\subsection{Mutation survival is not repository irreproducibility}
\label{sec:discussion-survival}

A SURVIVED mutation has a deliberately narrow meaning in our study. It means
that a confirmed non-equivalent mutation was not detected by the selected
frozen validation workflow. It does not imply that the repository is
irreproducible, that its published results are incorrect, or that no other
repository workflow would detect the same change.

This distinction is especially important because the selected workflows vary in
purpose. Some are upstream tests, whereas others are documented experiments,
examples, validation commands, or CI workflows. A repository may contain
additional checks that were not selected by the frozen protocol, and a
documented example may have been designed to demonstrate usage rather than to
validate numerical invariants.

Conversely, a KILLED mutation should not be interpreted as proof that a
repository is fully protected against reproducibility failures. A KILLED
outcome establishes only that the selected workflow rejected the particular
applied mutation under the study conditions.

Mutation testing is therefore used here as a probe of existing validation
behavior, not as a binary classifier of repository reproducibility.

\subsection{Workflow form and oracle strength}
\label{sec:discussion-oracles}

RQ2 did not reveal a monotonic relationship between the explicitness of the
validation oracle and mutation detection. In the prospectively classified B02
subset, neither of the two meaningful \texttt{assertion} cases was detected,
whereas two of thirteen \texttt{completion-only} cases were detected.

Taken alone, these proportions could be misleading. Both detected B02
mutations occurred in the same
\texttt{dependency-pin} $\times$ \texttt{completion-only} cell. No KILLED
mutation occurred in the other observed operator--oracle combinations.
Moreover, only two of seven selected \texttt{assertion} cases entered the
confirmed non-equivalent meaningful denominator, compared with thirteen of
twenty-two \texttt{completion-only} cases.

These observations prevent attributing the observed RQ2 pattern to oracle
strength. In particular, the study does not show that completion-only workflows
are stronger than assertion-based workflows.

The dependency result also illustrates why workflow completion can still act
as an effective detector for some classes of change. A dependency mutation can
alter the resolved environment in a way that causes imports, APIs, or runtime
behavior to fail before an explicit research-result assertion is reached. In
such a case, a completion-only workflow can kill the mutation even though it
contains no explicit numerical oracle.

This distinction suggests that workflow type, oracle type, and mutation type
capture different properties and should not be collapsed into a single notion
of ``test strength.''

\subsection{Executability as a prerequisite for mutation-based evaluation}
\label{sec:discussion-executability}

A second observation concerns executability. Only 13 of the 39 frozen
cases were evaluable during primary execution. Bounded restoration increased
this number to 24, leaving 15 cases non-evaluable.

These failures are methodologically distinct from mutation outcomes. A
repository that cannot be executed in the contemporary study environment
cannot be assigned a meaningful KILLED or SURVIVED result under the frozen
baseline-first protocol.

At the same time, the scale of the restoration problem is itself relevant to
empirical work on historical ML research software. We encountered barriers
associated with dependency resolution, obsolete package ecosystems, unavailable
or incompatible builds, removed APIs, undeclared runtime requirements, native
dependencies, specialized hardware assumptions, and unavailable workflow
requirements.

We do not interpret these barriers as direct evidence that the corresponding
research results are non-reproducible. They instead demonstrate that
retrospective execution can be constrained by software-ecosystem drift before
the scientific behavior of interest can even be evaluated.

The D027 layer substantially increased the evaluable sample while preserving
the frozen repository revision, candidate mutation, workflow, and oracle
classification. Separating restoration from the canonical primary result was
therefore important: it allowed us to recover additional empirical evidence
without rewriting the original execution record.

\subsection{Implications for ML research-software validation}
\label{sec:discussion-implications}

The results suggest that reproducibility-relevant validation may require checks
that are more specific than successful workflow completion alone. In
particular, workflows intended to guard experimental reproducibility can
benefit from making important experimental assumptions observable and
testable.

Examples include explicit checks on generated partitions, evaluation-protocol
parameters, expected dependency environments, or other artifacts that encode
experiment configuration. The appropriate safeguard depends on the scientific
workflow; the study does not imply that every repository should assert exact
numerical outputs or fix every dependency indefinitely.

A practical role for reproducibility-oriented mutation testing is therefore to
ask a targeted question: if a relevant experimental choice changes, does an
existing validation workflow notice?

This question complements conventional software testing. Traditional unit and
integration tests may establish functional correctness for APIs and components,
while reproducibility-oriented mutations probe whether validation is sensitive
to changes in experimental configuration that may still leave the software
operational.

For research-software developers, surviving confirmed non-equivalent mutations
can identify locations where existing validation does not currently constrain a
reproducibility-relevant choice. Such observations can motivate additional
checks where those choices are scientifically important, without implying that
every surviving mutation represents a defect.

\subsection{Role of MLReproMutate}
\label{sec:discussion-tool}

MLReproMutate was used in this study to make the mutation process explicit,
repeatable, and auditable. The empirical contribution is not merely the
application of mutation testing to ML software, but the use of a
reproducibility-oriented mutation model together with frozen repository
revisions, frozen validation workflows, explicit semantic-equivalence handling,
and structured execution evidence.

This separation between mutation generation, repository execution, semantic
verification, and analysis is important for future extensions. Additional
operators or repository populations can be studied without redefining the
meaning of the outcomes used here, while the current frozen corpus remains an
auditable empirical record.

\section{Threats to Validity}
\label{sec:threats}

\subsection{Construct validity}
\label{sec:threats-construct}

The four mutation operators represent a deliberately narrow subset of
reproducibility-relevant choices in ML research software. They do not cover all
sources of experimental variability, configuration drift, data-processing
choices, hardware effects, nondeterminism, or environment differences that may
affect reproducibility.

Accordingly, the mutations should be interpreted as controlled
reproducibility-relevant perturbations rather than as a complete model of real
ML faults. In particular, the study does not assume that every applied mutation
corresponds to a naturally occurring defect or that all reproducibility
failures can be represented by the four operators.

This limitation is partly intentional. The supported mutations were restricted
to mechanically identifiable syntax with explicit transformations, making
candidate selection and mutation application auditable. The resulting gain in
measurement precision comes at the cost of narrower construct coverage.

A second construct threat concerns the validation workflow used as the
detection oracle. The selected workflow represents one frozen repository
validation path, not the complete set of safeguards that may exist in a
repository. A SURVIVED mutation therefore means only that the selected workflow
did not detect the applied confirmed non-equivalent change. It does not imply
that no other test, experiment, manual inspection, or downstream scientific
analysis would detect it.

Similarly, a KILLED mutation establishes sensitivity to one applied mutation
but does not establish that the repository is generally protected against
reproducibility failures.

We mitigate these interpretation risks by separating workflow type from oracle
type, explicitly distinguishing completion-only workflows from workflows with
substantive acceptance conditions, and treating mutation outcomes as properties
of repository--mutation--workflow combinations rather than repository-level
reproducibility labels.

\subsection{Internal validity}
\label{sec:threats-internal}

A central internal-validity risk in mutation studies is outcome-dependent
selection. Repository choice, mutation-candidate choice, validation-workflow
choice, or oracle classification could otherwise be influenced by knowledge of
which cases are likely to produce KILLED or SURVIVED outcomes.

For the primary B02 corpus, we addressed this risk through static,
outcome-blind corpus construction. Repository eligibility, fixed revisions,
candidate mutations, validation workflows, and prospectively recorded oracle
categories were frozen before primary mutation execution. Candidate
repositories were not executed during corpus construction, and the corpus was
not extended or replaced after observing empirical outcomes.

The RQ2 classification was likewise separated from outcome analysis. Workflow
categories were inherited from the frozen corpus, and B02 oracle categories had
been recorded prospectively under protocol version 2. The RQ2 classification
frame was independently frozen before outcome joining. Consequently, the
observed RQ2 pattern was not used to redefine oracle categories.

Execution and restoration introduce another internal-validity concern. Changes
made solely to make historical software executable could alter the behavior
being measured. D027 therefore preserved the frozen repository revision,
mutation candidate, validation workflow, and oracle classification. Source
compatibility patches were prohibited, validation workflows could not be
weakened or skipped, restoration was baseline-first, and the restoration recipe
was fixed before evaluating the corresponding mutant. Baseline and mutant were
evaluated under symmetric restored conditions.

Nevertheless, restoration cannot eliminate all uncertainty associated with
executing historical software in a contemporary environment. A reconstructed
environment may differ from the original authors' environment in ways that are
not fully observable from repository metadata. We therefore retain the
canonical primary result separately and treat D027 as an additional bounded
evidence layer rather than as a replacement for primary execution.

Semantic-equivalence handling also affects internal validity. A surviving
syntactic mutation that does not change the relevant semantics should not
contribute evidence of weak validation. We therefore exclude confirmed
equivalent mutations from the meaningful detection denominator and distinguish
confirmed non-equivalent, confirmed equivalent, unverified, and not-run
semantic states.

\subsection{External validity}
\label{sec:threats-external}

The study does not constitute a random sample of all ML research repositories.
The corpus was constructed using operator-specific static eligibility criteria
and a fixed stopping rule. Consequently, the results should not be interpreted
as population-wide estimates of mutation detection or research-software
reproducibility.

The B02 operator frames are also unequal. The frozen stopping rule yielded ten
cases each for \texttt{random-seed} and \texttt{dependency-pin}, but the
eligible static frames for \texttt{data-split} and \texttt{cv-fold-count} were
exhausted at six and three cases. The meaningful evaluated denominators are
smaller still.

The repository population may also differ systematically from other forms of ML
software. The study focuses on research repositories for which both a supported
mutation candidate and a usable repository validation workflow could be
identified under the static protocol. Production ML systems, notebooks without
repository-level workflows, closed-source research code, and projects using
unsupported frameworks or configuration mechanisms are therefore outside the
observed frame.

These constraints limit generalization, but they also define the scope of the
empirical claim precisely: the study measures the behavior of the selected
frozen repository--operator cases under their selected validation workflows.

\subsection{Conclusion validity}
\label{sec:threats-conclusion}

The number of detected mutations is small. Across the 23 confirmed
non-equivalent evaluated mutations, only two were KILLED. RQ2 category sizes
are also sparse and uneven; for example, the prospective B02 oracle analysis
contains only two meaningful \texttt{assertion} cases.

We therefore treat RQ2 as descriptive and exploratory. We do not infer causal
effects of workflow type or oracle type, and we do not interpret differences in
observed percentages as evidence of statistical superiority.

This is especially important for the B02 oracle comparison. Both detected
mutations occurred in the
\texttt{dependency-pin} $\times$ \texttt{completion-only} cell. The observed
difference between \texttt{assertion} and \texttt{completion-only} workflows
can therefore reflect operator composition, sparse denominators, differential
evaluability, or other case-level factors rather than oracle strength.

For the same reason, operator-specific proportions such as the observed
detection among confirmed non-equivalent \texttt{dependency-pin} cases are
reported descriptively and should not be interpreted as estimates of a general
operator effect.

The primary quantitative claim is therefore intentionally narrow: in this
sample, the selected validation workflows detected 2 of 23 confirmed
non-equivalent reproducibility-relevant mutations.

\subsection{Temporal and executability validity}
\label{sec:threats-temporal}

The repositories in the corpus were executed after their frozen revisions had
been created, using contemporary software and infrastructure. Historical
dependency ecosystems, package indexes, compilers, operating-system libraries,
hardware availability, and external resources may have changed in the
intervening period.

This temporal distance was empirically consequential. Primary execution
evaluated only 13 of 39 frozen cases, and bounded restoration increased the
combined evaluated set to 24. The remaining non-evaluable cases therefore
reduce the observable empirical denominator and may introduce differential
attrition across workflow, operator, or oracle categories.

We do not interpret present-day execution failure as direct evidence that the
original research artifact was irreproducible at the time of publication.
Instead, these failures characterize the difficulty of retrospectively
executing the frozen software under the study conditions.

Conversely, successful contemporary restoration does not reconstruct the
original environment perfectly. D027 was designed to recover bounded
executability while preserving the frozen empirical target, not to claim
historical environment identity.

The study therefore reports primary and restored evidence separately before
combining successful evaluations at the analysis layer. This separation makes
the effect of temporal software-ecosystem drift visible rather than silently
removing non-evaluable cases from the study.

\section{Conclusion}
\label{sec:conclusion}

This study used mutation testing to examine whether validation workflows
already present in machine-learning research repositories detect controlled
changes in reproducibility-relevant experimental choices. Using MLReproMutate,
we evaluated mutations affecting random seeds, dependency constraints, data
splitting, and cross-validation configuration under an outcome-blind protocol
with frozen repository revisions, mutation candidates, and validation
workflows.

Of the 39 frozen repository--operator cases, 13 were evaluable during primary
execution. Bounded restoration increased the combined evaluable set to 24
cases. After excluding one confirmed-equivalent mutation, 23 confirmed
non-equivalent mutations entered the meaningful detection denominator. The
selected validation workflows detected 2 of these 23 mutations and did not
detect 21.

These results do not imply that the corresponding repositories are
irreproducible. Rather, they show that, for the evaluated cases in this sample,
the selected existing validation workflow did not detect the particular
reproducibility-relevant mutation in 21 of the 23 confirmed non-equivalent
evaluations.

The study also highlights executability as a practical prerequisite for
retrospective mutation analysis. Primary execution left 26 of the 39 frozen
cases non-evaluable; bounded restoration reduced this number to 15 while
preserving the primary execution record.

Our descriptive analysis did not support attributing mutation detection to
oracle explicitness. Both detected B02 mutations occurred in
\texttt{dependency-pin} cases with \texttt{completion-only} workflows, while
small and uneven category sizes, operator composition, and differential
evaluability prevent stronger inference.

More broadly, reproducibility-oriented mutation testing provides a way to ask
a concrete validation question: if an experimentally important choice changes,
does an existing repository safeguard notice? MLReproMutate provides an
auditable mechanism for posing this question through controlled mutations,
baseline-first execution, and explicit handling of semantic equivalence.

\section*{Software and Data Availability}

MLReproMutate version 0.1.0 and the frozen machine-readable empirical
artifacts accompanying this study are archived on Zenodo at
\url{https://doi.org/10.5281/zenodo.22126120}.
The development repository is available at
\url{https://github.com/ilyuka/MLReproMutate}.

\clearpage
\appendix

\section{Supplementary Empirical Accounting}
\label{app:supplementary}

The appendices report supporting accounting and diagnostic views of the frozen
empirical evidence. They do not introduce additional repository executions,
mutation outcomes, or post hoc case selection.

\subsection{Full frozen corpus}
\label{app:full-corpus}

Table~\ref{tab:full-corpus} lists all 39 frozen repository--operator cases used
in the empirical study. The table joins the final accounting layer with the
outcome-blind workflow frame by frozen case identifier. B01 oracle entries are
left unclassified because oracle kind was not prospectively recorded under the
schema used for that calibration batch.

The displayed outcome is the combined study result: the canonical primary
result when mutation evaluation succeeded at the primary layer, the bounded
D027 result when a previously non-evaluable frozen case was successfully
restored and evaluated, and non-evaluable otherwise. This presentation does
not alter the separate primary and restoration accounting reported in the main
text.

\begingroup
\scriptsize
\setlength{\tabcolsep}{2pt}
\renewcommand{\arraystretch}{1.15}
\begin{longtable}{@{}>{\raggedright\arraybackslash}p{0.075\textwidth}>{\raggedright\arraybackslash}p{0.245\textwidth}>{\raggedright\arraybackslash}p{0.115\textwidth}>{\raggedright\arraybackslash}p{0.145\textwidth}>{\raggedright\arraybackslash}p{0.110\textwidth}>{\raggedright\arraybackslash}p{0.070\textwidth}>{\raggedright\arraybackslash}p{0.110\textwidth}@{}}
\caption{Full frozen 39-case empirical corpus. Oracle classification is shown only where it was prospectively recorded for B02; B01 entries are left unclassified. Source identifies whether the combined result comes from the canonical primary execution or the bounded D027 restoration layer.}\label{tab:full-corpus}\\
\toprule
\textbf{Case} &
\textbf{Repository} &
\textbf{Operator} &
\textbf{Workflow} &
\textbf{Oracle} &
\textbf{Source} &
\textbf{Outcome} \\
\midrule
\endfirsthead

\multicolumn{7}{c}{\tablename\ \thetable\ --- continued from previous page} \\
\toprule
\textbf{Case} &
\textbf{Repository} &
\textbf{Operator} &
\textbf{Workflow} &
\textbf{Oracle} &
\textbf{Source} &
\textbf{Outcome} \\
\midrule
\endhead

\midrule
\multicolumn{7}{r}{Continued on next page} \\
\endfoot

\bottomrule
\endlastfoot

\texttt{B01-01} & \texttt{snu}\texttt{-}\allowbreak{}\texttt{causality}\texttt{-}\allowbreak{}\texttt{lab}\texttt{/}\allowbreak{}\texttt{efficient}\texttt{-}\allowbreak{}\texttt{canonical}\texttt{-}\allowbreak{}\texttt{bounding} & \texttt{dependency}\texttt{-}\allowbreak{}\texttt{pin} & \texttt{upstream}\texttt{-}\allowbreak{}\texttt{test} & -- & primary & \textsc{Survived} \\
\texttt{B01-02} & \texttt{scikit}\texttt{-}\allowbreak{}\texttt{learn}\texttt{-}\allowbreak{}\texttt{contrib}\texttt{/}\allowbreak{}\texttt{imbalanced}\texttt{-}\allowbreak{}\texttt{learn} & \texttt{data}\texttt{-}\allowbreak{}\texttt{split} & \texttt{documented}\texttt{-}\allowbreak{}\texttt{example} & -- & D027 & \textsc{Survived} \\
\texttt{B01-03} & \texttt{scikit}\texttt{-}\allowbreak{}\texttt{learn}\texttt{-}\allowbreak{}\texttt{contrib}\texttt{/}\allowbreak{}\texttt{MAPIE} & \texttt{random}\texttt{-}\allowbreak{}\texttt{seed} & \texttt{documented}\texttt{-}\allowbreak{}\texttt{example} & -- & primary & \textsc{Survived} \\
\texttt{B01-04} & \texttt{wwu}\texttt{-}\allowbreak{}\texttt{mmll}\texttt{/}\allowbreak{}\texttt{photonai} & \texttt{cv}\texttt{-}\allowbreak{}\texttt{fold}\texttt{-}\allowbreak{}\texttt{count} & \texttt{documented}\texttt{-}\allowbreak{}\texttt{example} & -- & primary & \textsc{Survived} \\
\texttt{B01-05} & \texttt{damianhorna}\texttt{/}\allowbreak{}\texttt{multi}\texttt{-}\allowbreak{}\texttt{imbalance} & \texttt{data}\texttt{-}\allowbreak{}\texttt{split} & \texttt{upstream}\texttt{-}\allowbreak{}\texttt{test} & -- & primary & \textsc{Survived} \\
\texttt{B01-06} & \texttt{KoheiObata}\texttt{/}\allowbreak{}\texttt{DMM} & \texttt{dependency}\texttt{-}\allowbreak{}\texttt{pin} & \texttt{documented}\texttt{-}\allowbreak{}\texttt{experiment} & -- & -- & non-\allowbreak{}evaluable \\
\texttt{B01-07} & \texttt{sigeisler}\texttt{/}\allowbreak{}\texttt{robustness}\texttt{\_}\allowbreak{}\texttt{of}\texttt{\_}\allowbreak{}\texttt{gnns}\texttt{\_}\allowbreak{}\texttt{at}\texttt{\_}\allowbreak{}\texttt{scale} & \texttt{dependency}\texttt{-}\allowbreak{}\texttt{pin} & \texttt{upstream}\texttt{-}\allowbreak{}\texttt{test} & -- & -- & non-\allowbreak{}evaluable \\
\texttt{B01-08} & \texttt{ContextLab}\texttt{/}\allowbreak{}\texttt{hypertools} & \texttt{random}\texttt{-}\allowbreak{}\texttt{seed} & \texttt{documented}\texttt{-}\allowbreak{}\texttt{example} & -- & primary & \textsc{Survived} \\
\texttt{B01-09} & \texttt{sherbold}\texttt{/}\allowbreak{}\texttt{autorank} & \texttt{random}\texttt{-}\allowbreak{}\texttt{seed} & \texttt{documented}\texttt{-}\allowbreak{}\texttt{example} & -- & primary & \textsc{Survived} \\
\texttt{B01-10} & \texttt{BorgwardtLab}\texttt{/}\allowbreak{}\texttt{P}\texttt{-}\allowbreak{}\texttt{WL} & \texttt{cv}\texttt{-}\allowbreak{}\texttt{fold}\texttt{-}\allowbreak{}\texttt{count} & \texttt{documented}\texttt{-}\allowbreak{}\texttt{experiment} & -- & primary & \textsc{Survived} \\
\texttt{B02-01} & \texttt{tslearn}\texttt{-}\allowbreak{}\texttt{team}\texttt{/}\allowbreak{}\texttt{tslearn} & \texttt{random}\texttt{-}\allowbreak{}\texttt{seed} & \texttt{documented}\texttt{-}\allowbreak{}\texttt{example} & \texttt{completion}\texttt{-}\allowbreak{}\texttt{only} & primary & \textsc{Survived} \\
\texttt{B02-02} & \texttt{tensorly}\texttt{/}\allowbreak{}\texttt{tensorly} & \texttt{random}\texttt{-}\allowbreak{}\texttt{seed} & \texttt{documented}\texttt{-}\allowbreak{}\texttt{example} & \texttt{completion}\texttt{-}\allowbreak{}\texttt{only} & primary & \textsc{Survived} \\
\texttt{B02-03} & \texttt{braindecode}\texttt{/}\allowbreak{}\texttt{braindecode} & \texttt{random}\texttt{-}\allowbreak{}\texttt{seed} & \texttt{documented}\texttt{-}\allowbreak{}\texttt{example} & \texttt{completion}\texttt{-}\allowbreak{}\texttt{only} & D027 & \textsc{Survived} \\
\texttt{B02-04} & \texttt{rtqichen}\texttt{/}\allowbreak{}\texttt{torchdiffeq} & \texttt{random}\texttt{-}\allowbreak{}\texttt{seed} & \texttt{documented}\texttt{-}\allowbreak{}\texttt{example} & \texttt{completion}\texttt{-}\allowbreak{}\texttt{only} & primary & \textsc{Survived} \\
\texttt{B02-05} & \texttt{EthanJamesLew}\texttt{/}\allowbreak{}\texttt{AutoKoopman} & \texttt{random}\texttt{-}\allowbreak{}\texttt{seed} & \texttt{documented}\texttt{-}\allowbreak{}\texttt{experiment} & \texttt{completion}\texttt{-}\allowbreak{}\texttt{only} & -- & non-\allowbreak{}evaluable \\
\texttt{B02-06} & \texttt{DistrictDataLabs}\texttt{/}\allowbreak{}\texttt{yellowbrick} & \texttt{random}\texttt{-}\allowbreak{}\texttt{seed} & \texttt{upstream}\texttt{-}\allowbreak{}\texttt{test} & \texttt{assertion} & -- & non-\allowbreak{}evaluable \\
\texttt{B02-07} & \texttt{kLabUM}\texttt{/}\allowbreak{}\texttt{rrcf} & \texttt{random}\texttt{-}\allowbreak{}\texttt{seed} & \texttt{upstream}\texttt{-}\allowbreak{}\texttt{test} & \texttt{assertion} & D027 & \textsc{Survived} \\
\texttt{B02-08} & \texttt{eric}\texttt{-}\allowbreak{}\texttt{mitchell}\texttt{/}\allowbreak{}\texttt{detect}\texttt{-}\allowbreak{}\texttt{gpt} & \texttt{random}\texttt{-}\allowbreak{}\texttt{seed} & \texttt{documented}\texttt{-}\allowbreak{}\texttt{experiment} & \texttt{completion}\texttt{-}\allowbreak{}\texttt{only} & -- & non-\allowbreak{}evaluable \\
\texttt{B02-09} & \texttt{gnina}\texttt{/}\allowbreak{}\texttt{libmolgrid} & \texttt{random}\texttt{-}\allowbreak{}\texttt{seed} & \texttt{upstream}\texttt{-}\allowbreak{}\texttt{test} & \texttt{assertion} & -- & non-\allowbreak{}evaluable \\
\texttt{B02-10} & \texttt{xucong}\texttt{-}\allowbreak{}\texttt{zhang}\texttt{/}\allowbreak{}\texttt{ETH}\texttt{-}\allowbreak{}\texttt{XGaze} & \texttt{random}\texttt{-}\allowbreak{}\texttt{seed} & \texttt{documented}\texttt{-}\allowbreak{}\texttt{experiment} & \texttt{completion}\texttt{-}\allowbreak{}\texttt{only} & -- & non-\allowbreak{}evaluable \\
\texttt{B02-11} & \texttt{princeton}\texttt{-}\allowbreak{}\texttt{nlp}\texttt{/}\allowbreak{}\texttt{SimCSE} & \texttt{dependency}\texttt{-}\allowbreak{}\texttt{pin} & \texttt{documented}\texttt{-}\allowbreak{}\texttt{validation} & \texttt{completion}\texttt{-}\allowbreak{}\texttt{only} & -- & non-\allowbreak{}evaluable \\
\texttt{B02-12} & \texttt{sunblaze}\texttt{-}\allowbreak{}\texttt{ucb}\texttt{/}\allowbreak{}\texttt{dpml}\texttt{-}\allowbreak{}\texttt{benchmark} & \texttt{dependency}\texttt{-}\allowbreak{}\texttt{pin} & \texttt{documented}\texttt{-}\allowbreak{}\texttt{experiment} & \texttt{completion}\texttt{-}\allowbreak{}\texttt{only} & -- & non-\allowbreak{}evaluable \\
\texttt{B02-13} & \texttt{rcamino}\texttt{/}\allowbreak{}\texttt{multi}\texttt{-}\allowbreak{}\texttt{categorical}\texttt{-}\allowbreak{}\texttt{gans} & \texttt{dependency}\texttt{-}\allowbreak{}\texttt{pin} & \texttt{documented}\texttt{-}\allowbreak{}\texttt{experiment} & \texttt{completion}\texttt{-}\allowbreak{}\texttt{only} & -- & non-\allowbreak{}evaluable \\
\texttt{B02-14} & \texttt{IBM}\texttt{/}\allowbreak{}\texttt{concept}\texttt{\_}\allowbreak{}\texttt{transformer} & \texttt{dependency}\texttt{-}\allowbreak{}\texttt{pin} & \texttt{upstream}\texttt{-}\allowbreak{}\texttt{test} & \texttt{assertion} & primary & \textsc{Equivalent} \\
\texttt{B02-15} & \texttt{apple}\texttt{/}\allowbreak{}\texttt{ml}\texttt{-}\allowbreak{}\texttt{cvpr2019}\texttt{-}\allowbreak{}\texttt{swd} & \texttt{dependency}\texttt{-}\allowbreak{}\texttt{pin} & \texttt{documented}\texttt{-}\allowbreak{}\texttt{example} & \texttt{completion}\texttt{-}\allowbreak{}\texttt{only} & D027 & \textsc{Killed} \\
\texttt{B02-16} & \texttt{GestureGeneration}\texttt{/}\allowbreak{}\texttt{Speech}\texttt{\_}\allowbreak{}\texttt{driven}\texttt{\_}\allowbreak{}\texttt{gesture}\texttt{\_}\allowbreak{}\texttt{generation}\texttt{\_}\allowbreak{}\texttt{with}\texttt{\_}\allowbreak{}\texttt{autoencoder} & \texttt{dependency}\texttt{-}\allowbreak{}\texttt{pin} & \texttt{documented}\texttt{-}\allowbreak{}\texttt{experiment} & \texttt{completion}\texttt{-}\allowbreak{}\texttt{only} & -- & non-\allowbreak{}evaluable \\
\texttt{B02-17} & \texttt{AaltoML}\texttt{/}\allowbreak{}\texttt{BayesNewton} & \texttt{dependency}\texttt{-}\allowbreak{}\texttt{pin} & \texttt{upstream}\texttt{-}\allowbreak{}\texttt{test} & \texttt{assertion} & -- & non-\allowbreak{}evaluable \\
\texttt{B02-18} & \texttt{pulimeng}\texttt{/}\allowbreak{}\texttt{eToxPred} & \texttt{dependency}\texttt{-}\allowbreak{}\texttt{pin} & \texttt{documented}\texttt{-}\allowbreak{}\texttt{validation} & \texttt{completion}\texttt{-}\allowbreak{}\texttt{only} & D027 & \textsc{Survived} \\
\texttt{B02-19} & \texttt{ddd9898}\texttt{/}\allowbreak{}\texttt{DeepNano} & \texttt{dependency}\texttt{-}\allowbreak{}\texttt{pin} & \texttt{documented}\texttt{-}\allowbreak{}\texttt{validation} & \texttt{completion}\texttt{-}\allowbreak{}\texttt{only} & primary & \textsc{Killed} \\
\texttt{B02-20} & \texttt{MKMaS}\texttt{-}\allowbreak{}\texttt{GUET}\texttt{/}\allowbreak{}\texttt{SSJE} & \texttt{dependency}\texttt{-}\allowbreak{}\texttt{pin} & \texttt{documented}\texttt{-}\allowbreak{}\texttt{experiment} & \texttt{completion}\texttt{-}\allowbreak{}\texttt{only} & -- & non-\allowbreak{}evaluable \\
\texttt{B02-21} & \texttt{thomas}\texttt{-}\allowbreak{}\texttt{young}\texttt{-}\allowbreak{}\texttt{2013}\texttt{/}\allowbreak{}\texttt{mindware} & \texttt{data}\texttt{-}\allowbreak{}\texttt{split} & \texttt{ci} & \texttt{completion}\texttt{-}\allowbreak{}\texttt{only} & D027 & \textsc{Survived} \\
\texttt{B02-22} & \texttt{Alcoholrithm}\texttt{/}\allowbreak{}\texttt{PTaRL} & \texttt{data}\texttt{-}\allowbreak{}\texttt{split} & \texttt{upstream}\texttt{-}\allowbreak{}\texttt{test} & \texttt{completion}\texttt{-}\allowbreak{}\texttt{only} & D027 & \textsc{Survived} \\
\texttt{B02-23} & \texttt{tfmortie}\texttt{/}\allowbreak{}\texttt{setvaluedprediction} & \texttt{data}\texttt{-}\allowbreak{}\texttt{split} & \texttt{upstream}\texttt{-}\allowbreak{}\texttt{test} & \texttt{completion}\texttt{-}\allowbreak{}\texttt{only} & D027 & \textsc{Survived} \\
\texttt{B02-24} & \texttt{lucazav}\texttt{/}\allowbreak{}\texttt{binclass}\texttt{-}\allowbreak{}\texttt{tools} & \texttt{data}\texttt{-}\allowbreak{}\texttt{split} & \texttt{upstream}\texttt{-}\allowbreak{}\texttt{test} & \texttt{assertion} & -- & non-\allowbreak{}evaluable \\
\texttt{B02-25} & \texttt{ECOLE}\texttt{-}\allowbreak{}\texttt{ITN}\texttt{/}\allowbreak{}\texttt{NguyenSSCI2021} & \texttt{data}\texttt{-}\allowbreak{}\texttt{split} & \texttt{upstream}\texttt{-}\allowbreak{}\texttt{test} & \texttt{completion}\texttt{-}\allowbreak{}\texttt{only} & -- & non-\allowbreak{}evaluable \\
\texttt{B02-26} & \texttt{tungtokyo1108}\texttt{/}\allowbreak{}\texttt{PolicySynth} & \texttt{data}\texttt{-}\allowbreak{}\texttt{split} & \texttt{documented}\texttt{-}\allowbreak{}\texttt{example} & \texttt{completion}\texttt{-}\allowbreak{}\texttt{only} & primary & \textsc{Survived} \\
\texttt{B02-27} & \texttt{rsteca}\texttt{/}\allowbreak{}\texttt{sklearn}\texttt{-}\allowbreak{}\texttt{deap} & \texttt{cv}\texttt{-}\allowbreak{}\texttt{fold}\texttt{-}\allowbreak{}\texttt{count} & \texttt{upstream}\texttt{-}\allowbreak{}\texttt{test} & \texttt{assertion} & D027 & \textsc{Survived} \\
\texttt{B02-28} & \texttt{ogozuacik}\texttt{/}\allowbreak{}\texttt{d3}\texttt{-}\allowbreak{}\texttt{discriminative}\texttt{-}\allowbreak{}\texttt{drift}\texttt{-}\allowbreak{}\texttt{detector}\texttt{-}\allowbreak{}\texttt{concept}\texttt{-}\allowbreak{}\texttt{drift} & \texttt{cv}\texttt{-}\allowbreak{}\texttt{fold}\texttt{-}\allowbreak{}\texttt{count} & \texttt{documented}\texttt{-}\allowbreak{}\texttt{experiment} & \texttt{completion}\texttt{-}\allowbreak{}\texttt{only} & D027 & \textsc{Survived} \\
\texttt{B02-29} & \texttt{farshidrayhancv}\texttt{/}\allowbreak{}\texttt{CUSBoost} & \texttt{cv}\texttt{-}\allowbreak{}\texttt{fold}\texttt{-}\allowbreak{}\texttt{count} & \texttt{documented}\texttt{-}\allowbreak{}\texttt{experiment} & \texttt{completion}\texttt{-}\allowbreak{}\texttt{only} & D027 & \textsc{Survived} \\
\end{longtable}
\endgroup

\subsection{Bounded restoration accounting}
\label{app:d027}

The primary empirical layer remains the canonical execution record. D027 was a
separate bounded restoration layer intended to recover evaluability for frozen
cases without changing the selected repository revision, mutation candidate,
validation workflow, or oracle.

Within the 24-case D027 cohort, 19 cases had a restoration report or assessment.
Substantive restoration was attempted for 18 cases. Eleven cases were restored
and reached mutation evaluation, seven were not restored after a substantive
attempt, and one additional reported case was not substantively attempted
because repository identity could not be recovered.

Table~\ref{tab:d027-accounting} reports this accounting separately from the
primary mutation outcomes.

\begin{table}[t]
\centering
\caption{D027 bounded-restoration accounting. Report presence and substantive restoration attempt are distinct.}
\label{tab:d027-accounting}
\small
\renewcommand{\arraystretch}{1.18}
\setlength{\tabcolsep}{7pt}
\begin{tabular}{lrr}
\toprule
\textbf{State} & \textbf{Count} & \textbf{\% of cohort} \\
\midrule
Cohort & 24 & 100.0\% \\
Report/assessment present & 19 & 79.2\% \\
No D027 report & 5 & 20.8\% \\
Substantive restoration attempted & 18 & 75.0\% \\
Report present, restoration not attempted & 1 & 4.2\% \\
Restored & 11 & 45.8\% \\
Not restored after substantive attempt & 7 & 29.2\% \\
Mutation evaluated & 11 & 45.8\% \\
\bottomrule
\end{tabular}
\end{table}

\subsection{Operator-by-oracle diagnostic}
\label{app:operator-oracle}

RQ2 was analyzed descriptively because the oracle categories were sparse and
operator composition was uneven. Table~\ref{tab:rq2-operator-oracle} reports
the B02 operator-by-oracle cross-classification used as a diagnostic for this
confounding.

In particular, both detected B02 mutations occurred in
\texttt{dependency-pin} cases classified as \texttt{completion-only}. The table
is therefore useful for showing why the observed oracle-level proportions
cannot be interpreted independently of mutation operator.

\begin{table}[t]
\centering
\caption{Operator composition of the prospective B02 oracle analysis. Both detected mutations occurred in the dependency-pin/completion-only cell; sparse cells prevent attribution of this pattern to oracle kind.}
\label{tab:rq2-operator-oracle}
\begin{tabular}{lrrrr}
\toprule
Operator & Oracle kind & N & Killed & Survived \\
\midrule
random-seed & assertion & 1 & 0 & 1 \\
random-seed & completion-only & 4 & 0 & 4 \\
dependency-pin & completion-only & 3 & 2 & 1 \\
data-split & completion-only & 4 & 0 & 4 \\
cv-fold-count & assertion & 1 & 0 & 1 \\
cv-fold-count & completion-only & 2 & 0 & 2 \\
\bottomrule
\end{tabular}
\end{table}

\FloatBarrier

\subsection{Evaluability by workflow and oracle category}
\label{app:rq2-attrition}

Mutation outcomes are observed only for cases that reach successful baseline
execution and mutation evaluation. Differential evaluability can therefore
change the composition of the analyzed subsets.

Table~\ref{tab:rq2-oracle-attrition} reports selected, evaluated, and meaningful case
counts by the categories used in RQ2. These counts are presented to make
attrition visible rather than implicitly conditioning the analysis on only
successful executions.

\begin{table}[t]
\centering
\caption{Evaluability of prospectively classified B02 oracle categories.}
\label{tab:rq2-oracle-attrition}
\begin{tabular}{lrrr}
\toprule
Oracle kind & Selected & Combined evaluated & Meaningful \\
\midrule
assertion & 7 & 3 & 2 \\
completion-only & 22 & 13 & 13 \\
\bottomrule
\end{tabular}
\end{table}

\FloatBarrier

\section{Mutation Operator Specification}
\label{app:operator-specification}

The empirical study used the concise labels
\texttt{random-seed},
\texttt{dependency-pin},
\texttt{data-split}, and
\texttt{cv-fold-count}. The corresponding implemented transformations were
deliberately narrow and deterministic once a mutation candidate had been
selected.

\begin{description}

\item[\texttt{random-seed}.]
Targets supported integer-literal seed calls and replaces the selected seed
\(N\) with \(N+1\). Supported calls in the study implementation include
Python's \texttt{random.seed}, NumPy seed calls, and
\texttt{torch.manual\_seed}.

\item[\texttt{dependency-pin}.]
Targets an exact dependency requirement of the form
\texttt{package==version} and replaces the exact constraint with
\texttt{package>=version}. Dependency evaluation independently resolves the
baseline and mutant environments. A textual mutation that resolves to the same
installed target version is treated as semantically equivalent.

\item[\texttt{data-split}.]
Targets a supported scikit-learn \texttt{train\_test\_split} call containing an
explicit non-\texttt{None} \texttt{stratify} argument and replaces the
stratification expression with \texttt{None}.

\item[\texttt{cv-fold-count}.]
Targets an explicit literal \texttt{n\_splits} value in a supported
KFold-family splitter and replaces \(N\) with \(N+1\). Supported splitters are
\texttt{KFold}, \texttt{StratifiedKFold}, \texttt{RepeatedKFold}, and
\texttt{RepeatedStratifiedKFold}.

\end{description}

Candidate detection and mutation application are separate operations.
Source-level candidates retain target-location metadata and are matched again
against the source before mutation application. This prevents a selected
candidate from being silently redirected to another syntactic occurrence.

\section{Outcome and Evidence Semantics}
\label{app:outcome-semantics}

The empirical analysis distinguishes execution state, mutation outcome, and
semantic verification.

A \textsc{Killed} mutation is one for which the frozen validation workflow
returns a failing status after a successful baseline. A \textsc{Survived}
mutation is one for which the same workflow succeeds after mutation.
Infrastructure, setup, baseline, and restoration failures are not mutation
outcomes.

Semantic verification is recorded independently. The relevant study-level
states are confirmed non-equivalent, confirmed equivalent, unverified, and not
run. Confirmed-equivalent mutations are excluded from the meaningful detection
denominator.

Consequently, the main detection measure is

\[
\frac{
  \textsc{Killed}
}{
  \textsc{Killed}
  +
  \text{confirmed non-equivalent \textsc{Survived}}
}.
\]

A surviving mutation therefore has a deliberately narrow interpretation: the
selected frozen validation workflow did not detect that particular confirmed
non-equivalent mutation. It is not a repository-level classification of
reproducibility.

\section{Empirical Provenance and Freeze Boundaries}
\label{app:freeze-boundaries}

The empirical study used explicit freeze boundaries to separate corpus
construction, execution, restoration, and analysis.

The 39-case empirical frame was frozen before post-freeze analysis. Candidate
repositories were not added, replaced, or rerun to improve the observed
mutation-detection rate. D027 restoration retained the frozen revision,
candidate, workflow, and oracle and did not rewrite the canonical primary
execution record.

For RQ2, workflow categories came from the frozen corpus metadata. Oracle
categories for the B02 cases were defined prospectively under protocol version
2 and frozen before joining them with mutation outcomes. The B01 calibration
cases predated this schema and were therefore not retrospectively assigned
oracle categories.

The repository contains the machine-readable case records, restoration
assessments, RQ2 blind frame, generated result tables, and manifests used to
produce the reported accounting. These artifacts are retained as the auditable
record underlying the manuscript tables.

% Bibliography
\bibliographystyle{plain}
\bibliography{references}

\end{document}